\documentclass{jps-cp}

\usepackage{txfonts} 
\newcommand{\Nmax}  	{N_{\mathrm{max}}}
\newcommand{\Nshell}  	{N_{\mathrm{shell}}}

\title{Perspectives on Nuclear Structure and Scattering
  with the {\it Ab Initio} No-Core Shell Model}

\author{James P. \textsc{Vary}$^{1}$,
  Pieter \textsc{Maris}$^{1}$,
  Patrick J. \textsc{Fasano}$^{2}$ and
  Mark A. \textsc{Caprio}$^{2}$}

\inst{
  $^{1}$Department of Physics and Astronomy, Iowa State University, Ames, Iowa  50011-3160, USA \\
  $^{2}$Department of Physics, University of Notre Dame, Notre Dame, Indiana  46556-5670, USA}

\email{jvary@iastate.edu}

\recdate{April 22, 2018}

\abst{Nuclear structure and reaction theory are undergoing a major
  renaissance with advances in many-body methods, strong interactions
  with greatly improved links to Quantum Chromodynamics (QCD), the
  advent of high performance computing, and improved computational
  algorithms. Predictive power, with well-quantified uncertainty, is
  emerging from non-perturbative approaches along with the potential
  for new discoveries such as predicting nuclear phenomena before they
  are measured. We present an overview of some recent developments and
  discuss challenges that lie ahead. Our focus is on explorations of
  alternative truncation schemes in the harmonic oscillator basis, of
  which our Japanese--United States collaborative work on the No-Core
  Monte-Carlo Shell Model is an example. Collaborations with Professor
  Takaharu Otsuka and his group have been instrumental in these
  developments.}

\kword{No-Core Shell Model, Monte-Carlo methods, Truncation methods, JISP16, Daejeon16}

\begin{document}
\maketitle

\section{Introduction}

With continuing advances in Leadership-Class computing and plans for
further developments leading to Exascale systems (defined as having
capabilities for $10^{18}$ floating-point operations per second
(flops)), computational scientists are developing quantum many-body
approaches that portend a new era of research and discovery in physics
as well as in other disciplines.  In particular, the nuclear physics
quantum many-body problem presents unique challenges that include the
need to simultaneously develop (1) strong inter-nucleon interactions
with ties to QCD in order to control the concomitant freedoms; (2)
non-perturbative many-body methods that respect all the underlying
symmetries; 
and (3) new algorithms
that prove efficient in solving the quantum many-body problem on
Leadership-Class supercomputers.  This triad of forefront requirements
impels multi-disciplinary collaborations that include theoretical
physicists, applied mathematicians and computer scientists.  These
requirements also foster international collaborations, such as the
Japan--United States collaboration, that can catalyze and incubate
new ideas while sharing the workload among the participating teams.

While the physics goals for computational nuclear structure and
reactions may seem obvious --- i.e., retaining predictive power and
quantifying the uncertainties --- the opportunities and challenges
presented with the continuing rapid development of supercomputer
architectures are less obvious to the broader community.  Simply put,
with the need to develop and apply fully microscopic approaches to
heavier nuclei as well as the need to include multi-nucleon
interactions and the coupling to the continuum, even Exascale
computers will be insufficient to meet all our plans. We therefore
must also work to develop truncation schemes that reduce the
computational burden without loss of fidelity to the underlying
theory.  In this work, we will focus on the second part of the triad
mentioned above --- the development of non-perturbative many-body
methods that respect the underlying symmetries.

\section{{\it Ab Initio} No-Core Monte Carlo Shell Model}

Several years ago, teams of theorists from Japan and the United States
initiated a joint research program aimed at benchmarking the No-Core
Monte Carlo Shell Model (NC-MCSM) with the No-Core Shell Model (NCSM)
using the same realistic interactions to define the nuclear
Hamiltonian.  This led to a series of projects
\cite{Abe:2011pk,Abe:2012wp,Abe:2013hva,Abe:2013x01,Abe:2013ypd,Abe:2013vqa,Abe:2014wja,Abe:2014x02}
that provided results for light nuclei while developing improved
methods for both approaches.  We investigated light nuclei up through
$A = 12$ with both approaches and compared their predictions for the
ground state energies.

In order to provide a perspective on this overview of benchmark
results as well as some exploratory work in the next section, we will
focus here on the similarities and differences in the Hamiltonian
basis spaces in the different approaches.  Both methods are applied
within a single-particle harmonic oscillator (HO) basis, and in both
methods the basis includes single-particle states up to a finite
number of HO shells designated by $\Nshell = 1 + 2\,n + l$, where $n$
is the radial quantum number and $l$ is the orbital angular momentum
quantum number.  Thus we count the $0s$ shell as the first shell.

With a finite number of single-particle states, the most general basis
for a many-body problem contains all possible many-body states
(configurations) that can be constructed from these single-particle
states, limited only by symmetry constraints.  Diagonalizing the
Hamiltonian in such a basis that contains all possible configurations
for a given single-particle basis is referred to as an Full
Configuration Interaction (FCI) calculation, and is considered the
'gold standard' in quantum chemistry.

However, the (naive) basis size $D$ of an FCI calculation grows
like
\begin{eqnarray*}
  D &=& \binom{N_{\mathrm{sp}}}{Z}\;\binom{N_{\mathrm{sp}}}{N}
\end{eqnarray*}
for an $A$-body calculation with $Z$ protons, $N$ neutrons, and
$N_{\mathrm{sp}}$ single-particle states.  With only four HO shells
($\Nshell=4$) we have $N_{\mathrm{sp}}=40$ and the FCI basis size
for $^{12}$C is of the order of $10^{16}$.  Even after applying
symmetry constraints, that will still be several orders of magnitude
beyond what can be diagonalized on current Leadership-Class computing
systems.  In addition, $\Nshell=4$ is not actually sufficient for
converging a calculation for $^{12}$C, as can be seen from
Fig.~\ref{fig:MCSMbenchmark} below.  Hence the need to further
truncate the many-body basis.

The NCSM uses a many-body basis truncation defined by $\Nmax$, the
number of HO quanta summed over all nucleons above the lowest possible
number of quanta for that nucleus~\cite{Barrett:2013nh}.  Such a
truncation scheme includes configurations with e.g. one nucleon in a
highly excited HO state and all others in the lowest HO states, as
well as configurations in which several nucleons (or even all nucleons
if $\Nmax \ge A$) are excited by one quantum.  When the NCSM results
are extrapolated to the infinite matrix limit, we refer to the results
as obtained in the No-Core Full Configuration (NCFC)
method~\cite{Maris:2008ax}.  In addition to drastically reducing the
basis size compared to an FCI calculation with the same highest
single-particle HO state, this particular truncation also leads to an
exact factorization of the center-of-mass motion and the intrinsic
motion of the self-bound
nuclei~\cite{Gloeckner:1974sst,Hagen:2009pq,Caprio:2012rv} --- and
ultimately, it is the intrinsic wavefunction that is required for
reaction calculations~\cite{Burrows:2017wqn,Gennari:2017yez}.

In the NC-MCSM, the many-body basis states are selected from the
underlying FCI basis defined by a single-particle truncation parameter
$\Nshell$.  It is a generalization of the Monte Carlo Shell Model
(MCSM), in which many-body states are constructed from linear
combinations of non-orthogonal angular-momentum and parity projected
deformed Slater determinants.  (For a review on the MCSM, see
Ref.~\cite{Otsuka:2001PPNP}.)  With increasing dimension of the Monte
Carlo basis space, the ground state energy of a NC-MCSM calculation
converges from above to the corresponding FCI value.  The energy,
therefore, always gives the variational upper bound for the exact
ground state energy.  Typically only a few hundred Monte Carlo basis
states are kept, though the underlying FCI basis can be of the order
of $10^{20}$.  If needed, the ground state energy (and other
observables) can be extrapolated by the energy variance
method~\cite{Abe:2012wp}.

\begin{figure}[t]
  \center\includegraphics[width=0.80\textwidth]{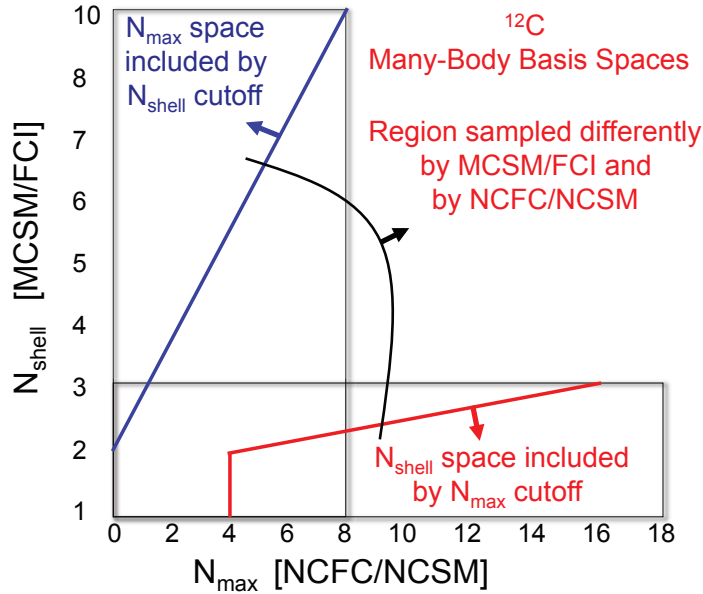}
  \caption{\label{fig:basis_space_coverage}
    (Color online) Overview of the bases covered with
    the NC-MCSM and NCSM methods for the case of $^{12}$C
    (adapted from Ref.~\cite{Abe:2012wp}).}
\end{figure}
Figure~\ref{fig:basis_space_coverage} sketches the topography of
many-body bases adopted in the different methods: the NC-MCSM (and
also the FCI method) and the NCSM method (with extrapolation for the
NCFC method).  This illustrates the different regions of the
single-particle space emphasized in the different approaches.  Of
course, the full (infinite dimensional) space is covered with
increasing either $\Nshell$ or $\Nmax$ and it becomes a practical
issue of the respective rates of convergence.

One of the advantages offered by the NC-MCSM approach is its
computational scaling with increasing number of nucleons $A$.  We
presented a study of the advantageous scaling properties of the
NC-MCSM in Ref.~\cite{Abe:2012wp} where we found that the rate of
increased demand on computational resources (for increasing $A$ at
fixed $\Nshell$) is orders of magnitude slower than for the NCSM (for
increasing $A$ at fixed $\Nmax$).  The question then turns to the
adequacy of extrapolation techniques for each method and the resulting
quantified uncertainties.  These issues have been addressed in our
subsequent efforts~\cite{Abe:2013ypd,Abe:2013vqa,Abe:2014wja,Abe:2014x02}.

Note that the rate of convergence depends on the observable so each
method is likely to have its advantages for certain observables.  One
may imagine that, qualitatively, the NC-MCSM/FCI approach is
advantageous for observables dominated by contributions from
multiparticle correlations in higher basis configurations while the
NCSM/NCFC favors observables that are sensitive to short-range
nucleon-nucleon ($NN$) correlations, though this is only a rough
picture.

\begin{figure}[t]
  \center\includegraphics[width=0.90\textwidth]{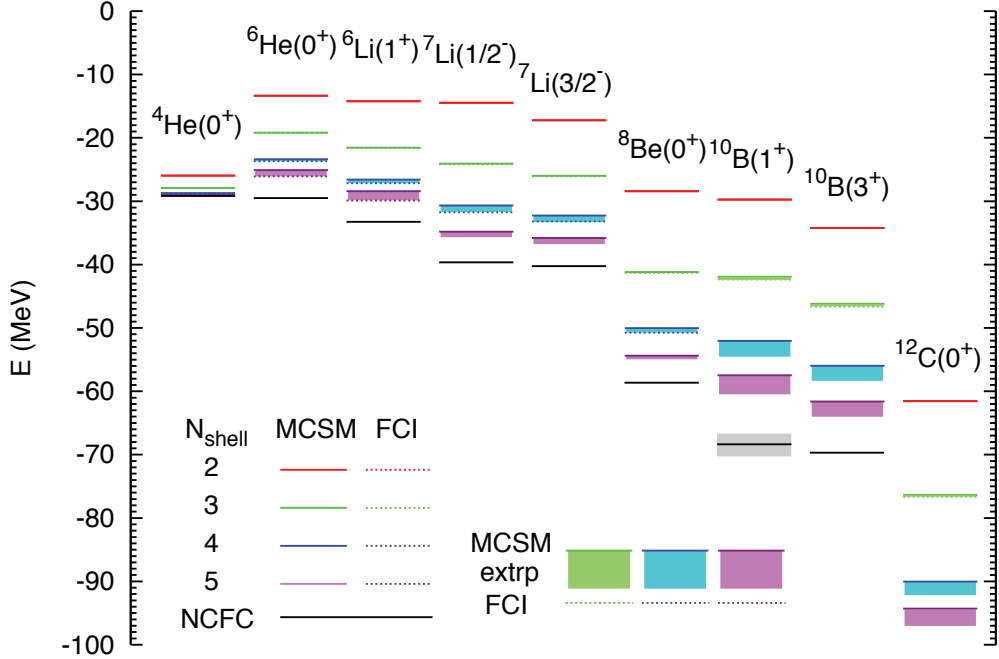}
  \caption{\label{fig:MCSMbenchmark}
    (Color online) Comparisons of the energies between the NC-MCSM
    and FCI along with the fully converged NCFC results where
    available. The NCFC result for the $^{10}$B($1^+$) state has a
    large uncertainty indicated by the grey band.  The NC-MCSM (FCI)
    results are shown as the solid (dotted) lines.  The extrapolated
    NC-MCSM results are illustrated by colored bands.  The NC-MCSM
    results, with extrapolation to their full $\Nshell$ basis
    \cite{Abe:2014wja}, nearly coincide with the FCI results.  From
    top to bottom, the truncation of the basis is $\Nshell$ = 2
    (red), 3 (green), 4 (blue), and 5 (purple).  Note that some
    results with $\Nshell$ = 4 and 5 were obtained only with the
    NC-MCSM (adapted from Ref.~\cite{Abe:2014wja}).  }
\end{figure}
In Fig.~\ref{fig:MCSMbenchmark}, we present recent
benchmarks of the ground state energies of several light nuclei using
the JISP16 $NN$ interaction~\cite{Shirokov:2005bk}.  We selected this
interaction since it produced a high-quality description of the $NN$
scattering data and was known to provide a good description of the
properties of light nuclei up to about
$A=12$~\cite{Maris:2013poa,Shirokov_review_JPV382:2014}.  For the
purposes of our benchmark we neglected the Coulomb interaction between
the protons.  We find that the NC-MCSM results with energy variance
extrapolation are nearly identical with the FCI results.  The
differences between the extrapolated NC-MCSM results and the NCFC
results provide a measure of the need for increasing $\Nshell$.
Fortunately, additional improvements to the NC-MCSM methods are under
development and larger $\Nshell$ values are already
achievable~\cite{Abe:2014wja}.

\begin{figure}[t]
  \center\includegraphics[width=0.65\textwidth]{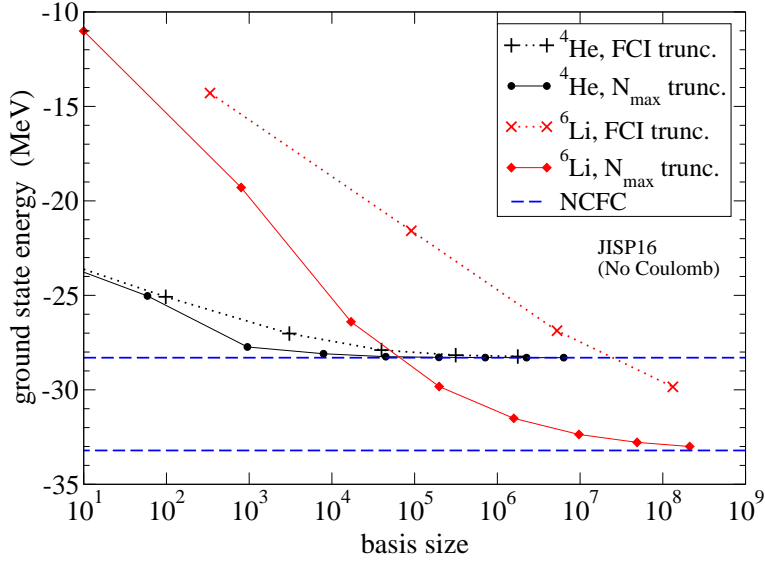}
  \caption{\label{fig_4He6Li_NmaxFCI}
    (Color online) Convergence of the ground state energy for $^4$He
    and $^6$Li with JISP16 (without the Coulomb interaction) as
    function of the many-body basis size.  NCSM results with $\Nmax$
    truncation are connected by solid lines.  FCI results with
    $\Nshell$ truncation are connected by dotted lines.  Both methods
    appear to converge from above, in accord with the variational
    principle, to the extrapolated NCFC result
    (adapted from Ref.~\cite{Maris:2012ee}).}
\end{figure}

In order to understand better the benchmark results, it is helpful to
have a comparison of convergence versus many-body basis size rather
than by comparing results directly between an FCI truncation and an
$\Nmax$ truncation.  In Fig.~\ref{fig_4He6Li_NmaxFCI}, we present
comparisons of the convergence rates for the ground states of $^4$He
and $^6$Li as a function of the many-body basis size using the FCI
truncation and the $\Nmax$ truncation schemes~\cite{Maris:2012ee}).
Both truncations approach the exact answer from above in concert with
the variational property of these approaches.  Clearly, the $\Nmax$
truncation provides faster convergence as a function of the
dimensionality.  One should keep in mind, however, the discussion
above concerning the very different computational scaling properties
with increasing $A$ of the NC-MCSM and the NCSM/NCFC approaches.  Good
computational scaling with increasing $A$ becomes overwhelmingly more
important at sufficiently large $A$.

\section{No-Core Shell Model with Alternative Truncation Schemes}

When we compare the FCI and the NCSM results above, we are comparing
different truncations of the many-body basis formed with HO
single-particle states.  It is natural to investigate whether
alternative truncation schemes are valuable in that they could provide
better converged results with comparable demand on computational
resources.  For example, we could examine alternatives to truncating
on the sum of the number of HO quanta in a many-body basis
configuration~\cite{Anderson:2012FB}.  Instead, we could also apply
specific weights to each single-particle orbital, characterized by
$n$, $l$, and $j$, and truncate the basis based on the sum of these
assigned weights in a many-body basis configuration.  In this
approach, the $\Nmax$ truncation of a conventional NCSM calculation is
recovered by choosing the weights to be $(2\,n+l)$.  Furthermore,
there is no reason for the underlying spatial single-particle
wavefunctions to be HO wavefunctions.  Indeed, alternative
single-particle basis functions such as the Laguerre
functions~\cite{Caprio:2012rv,Caprio:2014iha,McCoy:2016qyk} and the
natural orbital basis~\cite{Constantinou:2016urz} do both provide a
modest improvement of the convergence rates.

Here we explore possible improvements when we simply use
$W = (\alpha\, n + \beta\, l)$ but retain the HO single-particle
basis.  The coefficients $\alpha$ and $\beta$ define a weight $W$ for
each orbital and the many-body basis is truncated by a cutoff in the
sum over all nucleons of these weights for the orbitals in a many-body
basis configuration
\begin{eqnarray*}
  \sum_{i=1}^A  W_i \le W_{\max} \,.
\end{eqnarray*}
One can easily see that varying the coefficients $\alpha$ and $\beta$
allows a tradeoff between radial and orbital basis functions.  We will
examine only the ground state energies and root mean square (RMS)
point-proton radius here for a few selected cases for $\alpha$ and set
$\beta=1$ without loss of generality.  Similar convergence studies for
other single-particle basis functions, as well as other observables,
will be addressed in the future.  In addition, we could consider using
the total angular momentum $j$ in place of, or in addition to, the
orbital angular momentum $l$ in determining the weight, to give
different emphases to spin-orbit partner orbitals; and furthermore,
we could differentiate the weights for proton and neutron orbitals.

\begin{figure}[t]
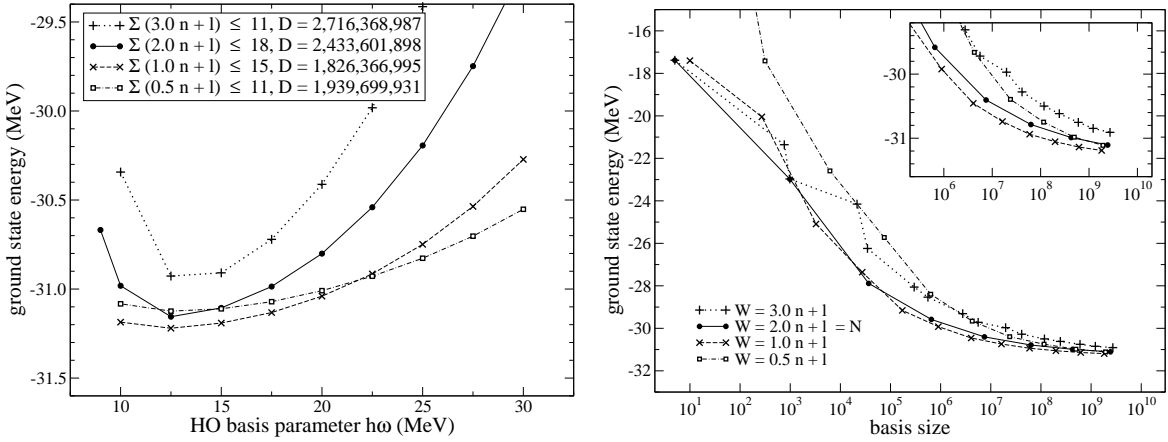

  \includegraphics[width=0.49\textwidth]{res_8He_Daejeon16_Egs.eps}\quad
  \includegraphics[width=0.484\textwidth]{res_8He_Daejeon16_Egs_hw15.eps}
  \caption{\label{fig_8He_Egs_gentrunc}
    Ground state energy of $^8$He with Daejeon16 as function of
    the HO basis parameter $\hbar\omega$ (left) and as function of the
    basis size at $\hbar\omega=15$~MeV (right) for using different
    single-particle weights $(\alpha \, n + l)$.}
\end{figure}
In Fig.~\ref{fig_8He_Egs_gentrunc} we present the ground state energy
of $^8$He with the Daejeon16 $NN$ interaction~\cite{Shirokov:2016ead}
at specific choices of $\alpha$ (with $\beta=1$).  The conventional
$\Nmax$ truncation is recovered for $\alpha=2$, in which case
$W_{\max} = 18$ corresponds to $\Nmax=14$ for $^8$He.  In order to
make these comparisons at approximately constant computational effort,
we select a cutoff $W_{\max}$ for each calculation (quoted in the
legend) such that it produces similar matrix dimensions of about two
billion.  We observe in the left panel of
Fig.~\ref{fig_8He_Egs_gentrunc} that, as we decrease the weight of
radial excitations compared to angular excitations, the dependence of
the ground state energy on the HO parameter $\hbar\omega$ decreases.
Furthermore, we see that $\alpha=1.0$, i.e. an orbital weight of
$W = (n + l)$, gives the lowest upper bound near the variational
minimum even though it has the smallest dimension among the cases
in this comparison set.

In the right panel of Fig.~\ref{fig_8He_Egs_gentrunc} we examine the
convergence rate of the ground state energy of $^8$He as a function of
the many-body basis size for the same set of choices for $\alpha$ as
in the left panel at a fixed value of the basis parameter
$\hbar\omega$.  As expected, all choices of weights produce
convergence from above with increasing basis size in concert with the
variational principle.  As the separate curves approach the same
asymptotic value, the differences among the curves may seem small on
this scale, but in the inset we can clearly see that for basis sizes
between a million and a few billion the calculations with $\alpha=1.0$
are closer to convergence than the other calculations with comparable
basis sizes.  For bases of 10 billion or more, $\alpha=0.5$ might
be an even better choice.

\begin{figure}[tbh]
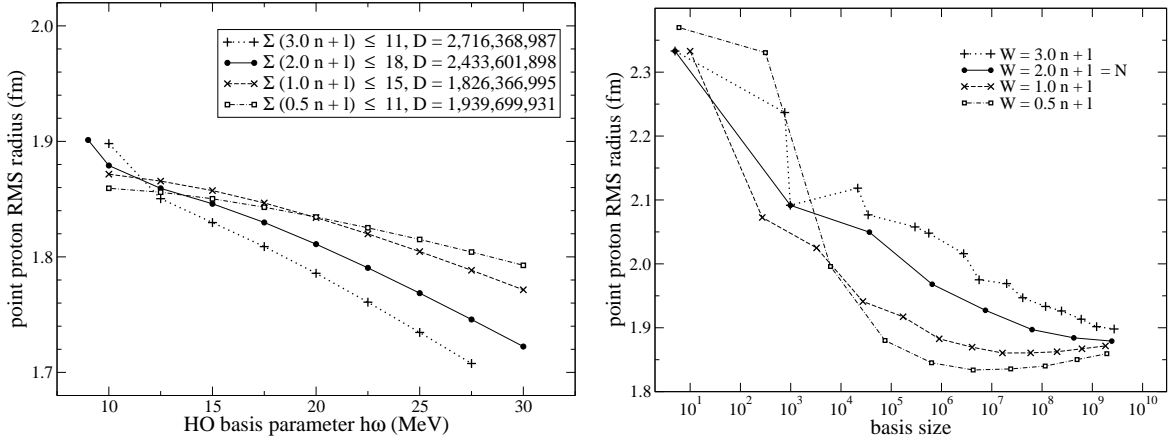

  \includegraphics[width=0.49\textwidth]{res_8He_Daejeon16_RMS.eps}\quad
  \includegraphics[width=0.484\textwidth]{res_8He_Daejeon16_RMS_hw10.eps}
  \caption{\label{fig_8He_RMS_gentrunc}
    Point-proton RMS radius of $^8$He with Daejeon16 as function of
    the HO basis parameter $\hbar\omega$ (left panel) and as function of the
    basis size at $\hbar\omega=10$~MeV (right panel) for different
    single-particle weights $(\alpha \, n + l)$.}
\end{figure}

Next, let us consider a different observable, namely the point-proton
RMS radius of $^8$He.  Note that this obervable is known to converge
slowly with $\Nmax$ in the conventional NCSM truncation, because the
$r^2$ operator is long-range and therefore sensitive to the asymptotic
tail of the wave function.  Our results are presented in
Fig.~\ref{fig_8He_RMS_gentrunc}, using these same weights for the
single-particle states and the same cutoffs in the many-body basis as
for the ground state energies.

First, we consider the dependence of the RMS point-proton radius on
$\hbar\omega$ in the left panel of Fig.~\ref{fig_8He_RMS_gentrunc}.
Again, as we decrease the weight of radial excitations compared to
angular excitations, the dependence of the RMS radius on the HO
parameter $\hbar\omega$ decreases.  However, without the benefit of
a variational principle for this observable, it is more challenging
to determine which truncation provides the most rapid convergence.
Nevertheless, greater independence of $\hbar\omega$ would be one
favorable indicator of improved convergence.  Among the cases
examined, the $(0.5\,n + l)$ case displays the smallest $\hbar\omega$
dependence over the entire $\hbar\omega$ range shown.  Furthermore,
the $\hbar\omega$ dependence is weakest in the low $\hbar\omega$ region where the
ground state energies are near their minima with respect to
$\hbar\omega$ as seen in the left panel of
Fig.~\ref{fig_8He_Egs_gentrunc} above.  Based on this criterion alone,
it would seem to suggest that the preferred weight for this observable
is $W = (0.5\,n + l)$ among the cases we examined, although that is
not the preferred weight obtained by considering the ground state
energy above.

Turning our attention to the right panel of
Fig.~\ref{fig_8He_RMS_gentrunc} we present the RMS point-proton radius
as a function of many-body basis size at a fixed value of the basis
parameter, that is $\hbar\omega=10$~MeV.  Here, the case $(n + l)$
provides the RMS radius results with the least sensitivity to
many-body basis size at higher dimensions.  Note that this is the
preferred weight for the convergence of the ground state energy.
Clearly, it would be worthwhile continuing these investigations to
larger many-body bases and different choices for the single-particle
weights in order to map out the convergence with respect to
the single-particle orbital weights and the many-body truncation.

\section{Conclusions and Outlook}

By benchmarking the NC-MCSM and the NCSM approaches, we confirm that
their respective extrapolated ground state energy results are in
agreement with expectations.  For the NC-MCSM, extrapolated ground
state energies using the energy variance method agree with the FCI
results where available.  Furthermore, the NC-MCSM and the NCSM
results for the ground state energy with increasing basis size are
consistent with each other.  However, the NC-MCSM results lie above
the NCSM results at comparable many-body basis sizes.  For
progressing to heavier nuclei, the NC-MCSM shows superior
computational scaling properties and is expected to provide valuable
{\it ab initio} results in these heavier systems where the NCSM has
limited utility at the present time.

We also explored different truncation schemes of the many-body basis
in the NCSM framework and found encouraging results by including more
harmonic oscillator single-particle states with higher radial quantum
numbers than would be included with the traditional $\Nmax$
truncation.
The ground state energy of $^8$He converged more rapidly with
increasing basis size using weights for the single-particle orbitals
of $W=(n+l)$, in combination with a cutoff $W_{\max}$ on the sum of
these weights for the many-body configurations in our basis.
Furthermore, both the ground state energy and the RMS point-proton
radius showed significantly improved independence of the basis
parameter $\hbar\omega$ with these single-particle weights.

There is much work to be done to more fully explore the opportunities
of alternative truncation schemes.  To list a few here, we mention:
(1) adopting natural orbitals or other single-particle bases;
(2) investigating additional nuclei; (3) adopting other Hamiltonians
including those with three-nucleon interactions; (4) mapping out
the convergence patterns of additional observables and (5)
developing and applying extrapolation methods for all observables.

We look forward to continuing our joint efforts with our colleagues in
Japan and we wish Takaharu Otsuka good health and many active and
enjoyable years ahead.

\section*{Acknowledgements}

We thank James Anderson for valuable discussions on alternative
truncation schemes, and Takaharu Otsuka for the hospitality during our
visits.  This material is based upon work supported by the
U.S.~Department of Energy, Office of Science, under Award
Numbers~DE-FG02-95ER-40934, DE-FG02-87ER40371, DE-FG02-06ER41407
(JUSTIPEN), DE-SC0018223 (SciDAC-4/NUCLEI) and DE-SC0015376 (DOE
Topical Collaboration in Nuclear Theory for Double-Beta Decay and
Fundamental Symmetries).  This research used computational resources
of the University of Notre Dame Center for Research Computing and of
the National Energy Research Scientific Computing Center (NERSC), a
U.S.~Department of Energy, Office of Science, user facility supported
under Contract~DE-AC02-05CH11231.


\end{document}